\documentclass[prb,nofootinbib,twocolumn,floatfix]{revtex4}
\usepackage{graphicx}
\graphicspath{{figures/}}
\usepackage{amsmath}
\bibpunct[, ]{[}{]}{,}{a}{,}{,}
\ifnum\lefthyphenmin<2\lefthyphenmin=2\fi
\ifnum\righthyphenmin<2\righthyphenmin=2\fi

\begin{document}
\title{Deposition of Atoms on a Quasicrystalline Substrate: Molecular-Dynamics Study in 3 Dimensions}
\author{Muhittin Mungan$^{1,2}$\footnote{ {\tt mmungan@boun.edu.tr}}, Yves Weisskopf$^{3}$, and  Mehmet Erbudak$^{1,3}$}
\affiliation{$^1$Department of Physics, Faculty of Arts and Sciences,\\
Bo\u gazi\c ci University, 34342 Bebek, Istanbul, Turkey,}
\affiliation{$^2$The Feza G\"ursey Institute,
P.O.B. 6, \c Cengelk\"oy, 34680 Istanbul, Turkey,}
\affiliation{$^3$Laboratory for Solid State Physics, ETH Zurich, 8093 Zurich, Switzerland}
\date{\today}

\begin{abstract}
We study the three-dimensional structure formation when atoms are deposited onto a substrate with a decagonal quasicrystalline order. Molecular-dynamics calculations show that the adsorbate layer consists of ordered nano-scale domains with orientations determined by the underlying substrate symmetry. Depending on the relative strength of the interactions of adsorbate atoms with each other and with the substrate atoms, different morphologies are observed ranging from layer-by-layer growth to cluster formation. We also find that the film thickness likewise affects the overall structure of the growing film: Depending on the relative strength of the interaction between adsorbate atoms, a structural transition of the configuration of the adsorbate layers closest to the substrate can occur as the number of deposited layers increases.  
\end{abstract}



\maketitle

Epitaxial growth produces structures with tailored properties, which do not exist in three dimensions (3D) due to thermodynamical restrictions. For example, it allows for the formation of nano-sized structures and thereby opens up new perspectives in science and technology. The large amount of experimental results  that have been obtained on self-size selecting growth over the last years poses an immense challenge for theorists. 

In this context, quasicrystalline media are of particular theoretical and experimental interest, because they are neither disordered like an amorphous solid nor periodic like a crystal. Contrary to common vitreous structures, quasicrystals do show perfect long-range order. Their surfaces can routinely be prepared, they are increasingly used as substrates for hetero-epitaxial growth and have been studied intensively over the last years \cite{Bolliger,Thomas,Ferralis,Rouven,curtar}. However, theoretical studies of this kind are complicated by the fact that the precise 3D structures of quasicrystals, i.e., their atomic configurations, are not readily available.  

Fl\"uckiger {\it et al.}~\cite{Thomas} have investigated the interface formed by an Al film grown on the surface of a decagonal Al$_{70}$Co$_{15}$Ni$_{15}$ (Al-Co-Ni) substrate and found that the mismatch at the interface causes the formation of well-oriented nanometric single crystals of Al, satisfying the epitaxial conditions on a local scale. Total energy calculations based on the simplifying assumption that the deposited layer interacts as a rigid lattice with the underlying substrate satisfactorily reproduced their experimental findings. In subsequent experimental work by Ferralis {\it et al.}~\cite{Ferralis}, the adsorption of Xe on the surface of decagonal Al-Co-Ni was investigated and results qualitatively similar to those found by Fl\"uckiger {\it et al.}~\cite{Thomas} were obtained. 

Recently, physisorption of Xe on decagonal Al-Co-Ni has been simulated using 3D Grand Canonical Monte Carlo techniques \cite{curtar,Setyawan, Setyawan_II}. Setyawan {\it et al.}  \cite{Setyawan} have shown that single locally hexagonal domains align only along fixed directions that have fivefold symmetry. However, since the size of the unit cell used is of the order of the domain size, it is difficult to observe the formation of multiple domains. This circumstance points towards the need of using much larger substrate sizes in the simulations that, additionally, will also reduce the effect of the sample boundaries.   

For the systems we consider, {\it i.e.}, adsorption of atoms with an essentially isotropic potential on a quasicrystalline substrate and not showing effects like intermixing, real-space experimental data on the formation of domains have been lacking. The dependence of the domain size and shape on experimental parameters is likewise not known. In a recent 2D study \cite{bilki} we have pointed towards the importance of the relative strength of the mutual interaction of adsorbate atoms (adatoms) compared to their interactions with the substrate atoms, which we quantified by the parameter $\eta$. We found that with increasing $\eta$ the adatoms assemble into well-formed and orientationally locked crystalline domains, qualitatively similar to the experimental findings~\cite{Thomas}.  However, the growth of adsorbate layers (adlayers) on a quasicrystal is a 3D process and should be treated as such.

In this article, we study the adlayer growth by carrying out extensive numerical simulations of the 3D deposition of 3.5 monolayers (ML) of 11 000 adatoms. As a substrate, we used 6 000 atoms in a surface bilayer of dimensions  $150\times150\times 2.05$ \AA $^3$ which was extracted from a model structure of the decagonal Al-Co-Ni quasicrystal recently proposed by Deloudi and Steurer \cite{Deloudi}. The large lateral size of the unit cell permits us to observe the formation of multiple domains and reduces artefacts generated by boundary conditions. We find that in addition to $\eta$, the film thickness also plays a role in determining the overall structure of the growing film, as one expects. The structure obtained after the formation of the first layer therefore does not necessarily determine the overall structure of the final film: As more and more layers are added to the film, the influence of the substrate on the first layer can diminish in favor of its natural structure. Thus, with increasing number of layers the first layer can undergo a structural transition. We also find that the parameter $\eta$ strongly influences the mode of growth:  Layer-by-layer growth at $\eta = 1$ gradually crossing over into 3D growth (clustering) at around  $\eta = 2$. For $\eta > 1.35$ we observe that the adatoms self-assemble into locally well-ordered domains with orientations determined by the symmetry of the underlying substrate, similar to what is observed in the  experiments \cite{Thomas, Ferralis}, and confirming our earlier 2D numerical modeling and simulations \cite{bilki}.

The adatoms are deposited by injecting them at an initial height $z_0$ above the substrate surface with a uniform initial velocity $v_0$. We assume that the substrate remains rigid and use a cooling mechanism to maintain the adlayer at a constant temperture $T$, as we will explain below. The interaction between adatoms as well as between adatoms and substrate atoms is assumed to be pairwise and of Lennard-Jones type, with the characteristic energy and length scales given by $\epsilon$ and $\sigma$, respectively.

Letting ${\bf r_\alpha}$ and ${\bf r_i}$ denote the coordinates of a substrate atom $\alpha$ and an adatom $i$, respectively, 
we numerically evaluate the equations of motion 
\begin{equation}
m \ddot{\bf r}_i = - \sum_{\alpha} {\bf \nabla} V(|{\bf r_i - r_\alpha}|) 
                   - \eta \sum_{j < i} {\bf \nabla} V(|{\bf r_j - r_i}|) + {\bf f}_i(T), 
\label{eom}
\end{equation}
where the parameter $\eta$ is used to adjust the relative strength of the adatom interactions with respect to the substrate. 

The adatoms are maintained at a constant temperature $T$ by implementing a thermostat proposed by Gilmore and Sprague~\cite{Gilmore}: Excess kinetic energy of the adatoms is dissipated by an additional friction force ${\bf f}_i(T)$ that is given by the product of a damping constant $\gamma$ times the excess kinetic energy with respect to the desired average kinetic energy.  

\begin{figure}[!b]
\includegraphics[width=.74\columnwidth]{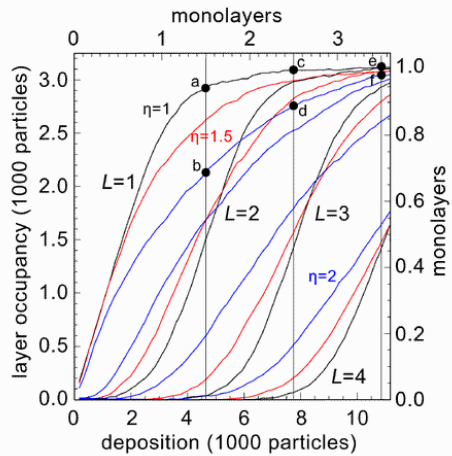} 
\caption{(color online) Layer occupancies in layers $L=1-4$ as a function of the total number of adatoms deposited. Shown are results  for relative interaction strengths $\eta = 1$, 1.5, and 2.}
\label{fig:depo}
\end{figure}

The numerical simulations are carried out in a unit cell of $150\times150\times50$ \AA{}$^{3}$ with periodic boundary conditions in the lateral ($x,y$) directions and a hard ceiling at $z = 50$ \AA. The quasicrystalline substrate is taken as a bilayer with a layer spacing of $2.05$ \AA\ containing about 3\,000 atoms in each layer. It is assumed that the substrate is sufficiently large and the effects of artificial periodicity induced by the boundary conditions can be neglected. For the interaction  parameters we used $\sigma = 2.55$~\AA\ corresponding to the lattice parameter of Al metal, while  $\epsilon$ is set to $0.25$ eV for practical purposes.  

Adatoms were inserted in batches of $180$ from a height of $z_0 = 25$ with an initial velocity of  $v_0 = 5$ (in time units where the mass of the adatoms is set to unity). In order to avoid correlations between the incident adatoms they were injected at random lateral $(x,y)$ positions, such that the distance between any two of them was at least $3 \sigma$. The substrate was maintained at a temperature $kT = 0.1$ eV, and a damping coefficient  $\gamma = 0.25$ was used. It was determined in advance that a waiting time of $t_{\rm wait}  = 25$ time units was sufficient for the incident adatoms to dissipate their energy to the adlayer, and for the latter in turn to relax back to the thermostat temperature $T$. Thus, in our simulations we inject $180$ adatoms at a time, wait for $t_{\rm wait}$ before inserting another batch. The equations of motion, Eqs.~(\ref{eom}), were integrated using a fourth-order Runge-Kutta method with adaptive step size \cite{NumRec}. 

We first investigate the occupancy of the adlayers as a function of the total number of adatoms deposited onto the substrate. For $\eta$ values of 1, 1.5, and 2, adlayer structures of 3.5 ML containing 10980 adatoms were grown, as described above. For each $\eta$ the individual layer heights were determined from the height distributions of the adatoms. The results are shown in Fig.~\ref{fig:depo}, where each continuous curve consists of  61 data points, corresponding to the 61 batches of 180 adatoms injected upon the surface.  We see that for $\eta$ values close to one the deposition curves are rather steep with a slope of nearly one, reflecting a layer-by-layer type of growth. For larger $\eta$ values the characteristic deposition curves  are flatter, implying that multiple layers are grown simultaneously. 

\begin{figure}[h]
\includegraphics[width=.55\columnwidth]{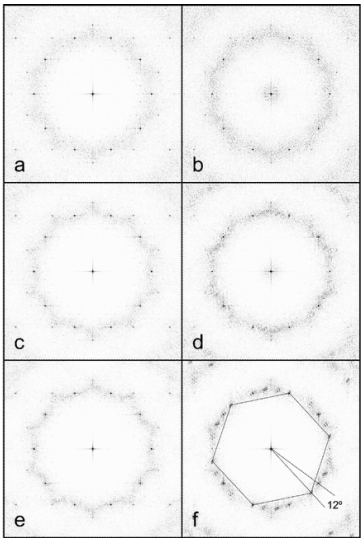}
\caption{Intensity patterns obtained from the Fourier transformation of the adatom structure of the first layer for $\eta = 1$ (left column) and $\eta = 2$ (right column) for the deposition of 1.5, 2.5, and 3.5~ML of adatoms (from top to bottom). In (f), one set of six spots is marked by a hexagon to emphasize the sixfold symmetry of the domain that gives rise to the pattern. The azimuthal orientation of the next domain is 12$^{\circ}$ away.}
\label{fig:thick}
\end{figure}

In order to investigate the influence of the substrate structure on the growing film as a function of film thickness, we follow the structure of the adlayer closest to the substrate. Fig.~\ref{fig:thick} displays a set of intensity patterns obtained from the Fourier transformation of the adatom structure of the first layer for $\eta = 1$ (left column) and $\eta = 2$ (right column) for the deposition of 1.5, 2.5, and 3.5~ML of adatoms (from top to bottom). Note that for $\eta = 1$  the interface layer of the growing film shows a tenfold symmetry, induced by the quasicrystalline substrate, that does not change appreciably with increasing film thickness.  This is in contrast to the growth mode at $\eta = 2$, where the structure of the film at the interface changes from quasicrystalline to local hexagonal domains with orientations locked to the substrate symmetry. One sixfold-intensity pattern is highlighted by the hexagon in Fig.~\ref{fig:thick} (f). The layer thicknesses have been indicated in Fig.~\ref{fig:depo} as thin vertical lines. The layer occupancies corresponding to the patterns shown in Figs.~\ref{fig:thick} (a) -- (f) have also been marked in Fig.~\ref{fig:depo}.

\begin{figure}
\includegraphics[width=.63\columnwidth]{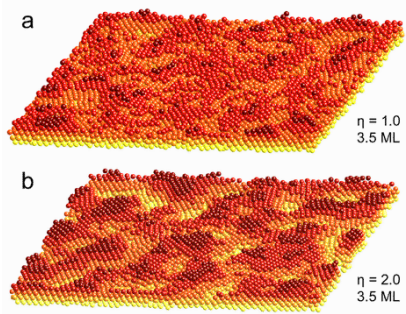}
\caption{(color online) Overall adatom configuration at 3.5~ML coverage for $\eta = 1$ (a) and $\eta = 2$ (b). Atoms belonging to a particular layer have the same color. The interface layer ($L=1$) is yellow. }
\label{fig:adconf}
\end{figure}

\begin{figure}[h]

\includegraphics[width=.679\columnwidth]{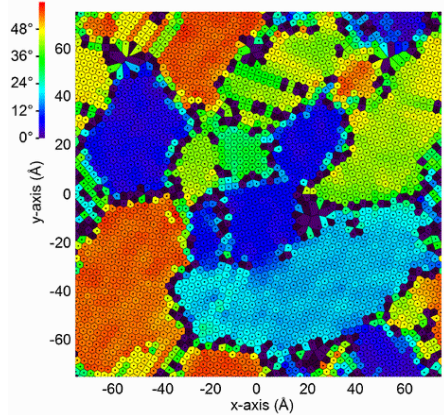}

\caption{(color online) Voronoi tessalation of the real-space configuration of adatoms (black spots) in the interface layer. Hexagons have been color coded with respect to their orientation relative to the $x$-axis. Voronoi $n$-gons with $n \ge 7$ and $n \le 5$ have been colored in light and dark violet, respectively, in order to clearly mark non-hexagonal polygons.}
\label{fig:voronoi}
\end{figure}

We now turn to the real-space configuration of the film. Figs.~\ref{fig:adconf} (a) and  (b) present the overall atomic configuration of 3.5~ML film thickness for $\eta = 1$ and $\eta = 2$, respectively. It is  evident that $\eta = 1$ results in a layer-by-layer growth with relatively high occupancies of the deeper layers, while the film produced with $\eta = 2$ is highly corrugated. These results are consistent with the deposition curves presented in Fig.~\ref{fig:depo}. The $\eta = 2$ film displays well-formed hexagonal domain structure. In a previous work, we have used Voronoi tessalation to reveal the domains of hexagonal structure in a 2D configuration of adatoms \cite{bilki}. We apply this technique to the configuration of adatoms in the interface layer calculated with $\eta = 2$, Fig.~\ref{fig:adconf} (b). The result is shown in Fig.~\ref{fig:voronoi}. Voronoi polygons other than hexagons are colored dark (light violet for $n \ge 7$-gons and dark violet for $n \le 5$-gons), while hexagons have been colored according to their orientation relative to the $x$-axis as indicated by the color-coded scale from $0-60^{\circ}$ on the left-hand side  of the figure. We distinguish at least four well-oriented hexagonal domains of orientations $6^{\circ}$ (dark blue), $18^{\circ}$ (light blue), $42^{\circ}$ (light green), and $54^{\circ}$ (orange), consistent with the corresponding Fourier intensity pattern in Fig.~\ref{fig:thick} (f). The domains have sharp boundaries consisting of non-hexagonal polygons, {\it i.e.}, topological defects. We also see some continuous transitions between domains with orientations differing by $12^{\circ}$.  Notice that for larger differences of orientations of neighboring domains the domain boundaries seem to consist mostly of topological defects.  Also note the large domain sizes of the order of 50~\AA\ in Fig.~\ref{fig:voronoi}. This shows directly the necessity of using appreciably large substrates in order to accomodate the formation and interactions of multiple domains while reducing the effects of boundary conditions.

Our 3D results on a hetero-epitaxial system consiting of a quasicrystal and an ordinary crystal show that the structural misfit can lead to the formation of nano-scale crystalline domains. The aperiodic substrate further acts as a structural template to azimuthally align these domains along the distinct symmetry directions of the quasicrystal. Moreover, we find that the domain formation and structure within a given adatom layer is consistent with our earlier findings on a 2D system in which single layers of adatoms were confined to a plane \cite{bilki}. These results are also in agreement with experimental findings~\cite{Thomas,Ferralis,curtar}. 

The clustering growth resulting in a well-formed local crystalline structure at $\eta = 2$, as clearly seen in Fig.~\ref{fig:adconf} (b), shows that the formation of the growing film is dominated by the relatively strong adatom-adatom interactions in comparison with the adatom-substrate interaction which causes the injected adatoms to preferentially coagulate with already present and cooled adatoms forming nano-clusters that serve as seeds. This is in contrast to the wetting behavior of the growing film at $\eta = 1$, shown in Fig.~\ref{fig:adconf} (a). Hence, by varying the parameter $\eta$ it is possible to continuously cover a wide range of growth modes.  For $\eta = 2$, we also observe a structural transition in the interface layer of the film as it gradually grows. The presence of multiple adlayers on top of the interface layer shows a strong tendency to revert the latter to its natural sixfold orientational symmetry. We believe that this transition is caused by the structural stiffening of the film with increasing number of adlayers. 

Experimental results of Fl\"uckiger {\it et al.} \cite{Thomas} for Al films grown on Al-Co-Ni show a transition from quasicrystalline structure at 1 ML to a structure of well-oriented nanoscale single crystals of Al at coverages of 2 ML and above (see also Ref. \cite{Ferralis} for similar results of Xe adsorbed on Al-Co-Ni). We observe qualitatively similar results for $\eta = 2$. In fact, additional simulations not presented here show that such a transition persists for values of $\eta$ as low as 1.35. It is very likely that such a threshold value depends also on the damping coefficient which determines the rate at which arriving adatoms cool down on the adlayer. A lower damping constant will increase the relaxation times for these atoms and could thus permit the exploration of a larger set of energetically more favorable local configurations. Although the layer-by-layer growth seen for $\eta \sim 1$ suggests a less pronounced dependency in this regime, it is possible that 
the structure of the wetting adlayer might nevertheless be affected by larger thermal relaxation times. We observed in a previous simulated annealing study of a single adlayer a dependence of the local structure on cooling rate \cite{bilki}.  

The present article considers the formation of structure on a quasicrystalline substrate in its dependence on the relative strength of adatom interactions and film thickness. A study investigating the effect of substrate temperature as well as the thermal relaxation rate is currently in progress. 

The authors gratefully acknowledge Sofia Deloudi for providing us with the coordinates of the Al-Co-Ni surface prior to publication. The computations were mainly done using the computer cluster Gilgamesh at the Feza G\"ursey Institute. This work has been funded in part by grant 04B303 of Bo\u gazi\c ci University. Financial support by Schweizerischer Nationalfonds is appreciated.


\begin{thebibliography}{99}
\newpage
\bibitem{Bolliger}B. Bolliger, V.E. Dmitrienko, M. Erbudak, R. L\"uscher, H.-U. Nissen, and A.R. Kortan, Phys. Rev. B {\bf 63}, 52203 (2001); V. Fourn\'ee and P.A. Thiel, J. Phys. D: Appl. Phys. R {\bf 83}, 38 (2005); P. Moras,  Y. Weisskopf, J.-N. Longchamp, M. Erbudak, P.H. Zhou, L. Ferrari, and C. Carbone, Phys. Rev. B {\bf 74}, 121405(R) (2006).

\bibitem{Thomas} T. Fl\" uckiger, Y. Weisskopf, M. Erbudak, R. L\" uscher, and A.R. Kortan, Nano Lett. {\bf 3}, 1717 (2003).

\bibitem{Ferralis} N. Ferralis, R.D. Diehl, K. Pussi, M. Lindroos, I. Fisher, and C.J. Jenks, 
Phys. Rev. B {\bf 69}, 075410 (2004).

\bibitem{Rouven}R. L\"uscher, M. Erbudak, and Y. Weisskopf, Surf. Sci. {\bf 569}, 163 (2004).

\bibitem{curtar}S. Curtarolo, W. Setyawan, N. Ferralis, R.D. Diehl, and M.W. Cole, Phys. Rev. Lett. {\bf 95}, 136104 (2005). 

\bibitem{Setyawan} W. Setyawan, N. Ferralis, R.D. Diehl, M.W. Cole, and S. Curtarolo, 
Phys. Rev. B {\bf 74}, 125425 (2006).

\bibitem{Setyawan_II} W. Setyawan, R.D. Diehl, N. Ferralis, M.W. Cole, and S. Curtarolo, 
J. Phys.: Condens. Matter {\bf 19}, 016007 (2007).

\bibitem{bilki}B. Bilki, M. Erbudak, M. Mungan, and Y. Weisskopf, Phys. Rev. B {\bf 75}, 045437 (2007).

\bibitem{Deloudi} S. Deloudi and W. Steurer,  Phil. Mag., in print.

\bibitem{Gilmore} C.M. Gilmore and J.A. Sprague, Phys. Rev. B {\bf 44}, 8950 (1991).

\bibitem{NumRec} {\it Numerical Recipes}, W.H. Press, B.P. Flannery, S.A. Teukolsky, and W.T. Vetterling, 2nd edition, Cambridge University Press, Cambridge, 1993.

\end{thebibliography}
\end{document}